# Microsecond electro-optic switching of nematic liquid crystals with giant dielectric anisotropy

Bing-Xiang Li [1,2], Greta Babakhanova [1,2], Rui-Lin Xiao [1,2], Volodymyr Borshch [1,2], Simon Siemianowski [3], Sergij V. Shiyanovskii [1,2], and Oleg D. Lavrentovich [1,2,4*]

[1]Advanced Materials and Liquid Crystal Institute, Kent State University, Kent, Ohio, 44242, USA
[2]Chemical Physics Interdisciplinary Program, Kent State University, Kent, Ohio, 44242, USA
[3]Merck KGaA, Darmstadt, Germany, 64293, Germany
[4]Department of Physics, Kent State University, Kent, OH, 44242, USA
*olavrent@kent.edu

Nematic liquid crystals exhibit a fast optical response when the applied electric field modifies the degree of order but does not change the direction of molecular orientation. The effect requires a relatively high electric field, on the order of $10^8$ V/m for a field induced birefringence change of 0.01. To address this detrimental issue, this work explores electrically induced modification of the order parameter in a material with a giant dielectric anisotropy of +200. A relatively weak field $3 \times 10^7$ V/m causes a significant change of birefringence, by about 0.04. Both switching on and off times are ~10 µs. The effect is called a microsecond electrically modified order parameter (MEMOP) and can be used in electro-optical devices, such as fast electro-optic shutters, phase modulators, and beam-steerers that require microsecond response times.

## I. INTRODUCTION

Nematic liquid crystals (NLCs) are used extensively in electro-optic applications because of their anisotropic dielectric and optical properties [1-3]. The average orientation direction of NLC molecules, called a director $\hat{\mathbf{n}}$, is also the optic axis. Most of the electro-optical applications of NLCs exploit the so-called Frederiks transition, [4] i.e., reorientation of $\hat{\mathbf{n}}$ by an external electric field followed up by elastic relaxation of $\hat{\mathbf{n}}$ when the field is switched off.

The latter process is very slow, as the driving force is a surface anchoring of $\hat{\mathbf{n}}$ at the bounding plates. The time of director reorientation is determined by the ratio of the viscous and elastic coefficients of the NLC, which yields a response time on the scale of milliseconds and above. Many research groups have explored ways to reduce the switching time to the millisecond or sub-millisecond range by optimizing the dielectric and viscoelastic parameter of NLCs, using dual frequency NLCs [5-7], introducing polymer-layer-free alignment [8], and designing the spatial configuration of electric fields [9]. Still, the response time of the typical nematic-based electro-optic devices in which the electric field realigns the director $\hat{\mathbf{n}}$, is rarely reaching the scale of 100 µs or less.

Recently, an alternative approach to electro-optic switching of NLCs has been proposed [10-12]. In this approach, the electric field modifies the degree of molecular order but does not realign $\hat{\mathbf{n}}$. The effect, called the nanosecond electrically modified order parameter (NEMOP), has been demonstrated for NLCs with a negative dielectric anisotropy [10-12]. In these materials, with a typical $\Delta\varepsilon \approx -10$, the field-induced birefringence, $\delta n$, is about 0.02 for an applied electric field on the order of $10^8$ V/m, while the switching speed is tens of nanoseconds. The current challenge is to reduce the driving field as compared to the NEMOP effect, while enhancing the birefringence change and making the response time significantly shorter than in the conventional Frederiks effect. In this work, we demonstrate that by using an NLC composition abbreviated as GPDA200 (Merck KGaA), with a giant positive dielectric anisotropy (GPDA), $\Delta\varepsilon > +200$, one can achieve a significant birefringence change of 0.04 at the applied field $3\times10^7$ V/m with the switching times on the order of 10 µs for both field-on and -off switching. We call the effect a microsecond electrically modified order parameter, or MEMOP.

## II. MATERIALS AND METHODS

### A. Dielectric and optical properties of the NLC

To measure the birefringence of the the mixture GPDA200, we placed a planar cell of thickness $d = 10.8 \, \mu\text{m}$ between two crossed polarizers in such a way that the rubbing direction of the cell is at the angle of 45 degrees with respect to the polarizers. An AC voltage is applied using transparend indium tin oxide (ITO) electrodes at the glass bounding plates of the cell, to cause reorientation of $\hat{\mathbf{n}}$ towards the field direction. The birefringence $\Delta n$ measured with a probing beam of He-Ne laser of wavelength $\lambda = 632.8 \, \text{nm}$, is calculated as from the field dependency of the transmitted intentity,

$$\Delta n = \frac{\lambda}{2d} \left( j + \frac{1}{2} + \frac{(-1)^j}{\pi} \arcsin \frac{2I - I_{max} - I_{min}}{I_{max} - I_{min}} \right), \quad (1)$$

where $j$ shows how many times the transmitted light intensity oscillates between the maximum intensity $I_{max}$ and the minimum intensity $I_{min}$, $I$ is the transmitted intensity in absence of the field, Fig.1(a). For dielectric measurements, we used an LCR meter (HP4284A) and home-made cells of thicknesses $d = 19.1 - 19.9 \, \mu\text{m}$ with planar and homeotropic anchoring. The planar alignment was achieved with antiparallel rubbed PI2555 polyimide layers (HD Microsystems), while the homeotropic alignment was realized using an SE1211 polyimide layer (Nissan Chemical Industries). The dielectric permittivity $\varepsilon_\parallel$ along $\hat{\mathbf{n}}$ is huge and highly dependent on both the temperature and field frequency, following the Debye model with a single relaxation time $\tau_D$, Fig.1(b). In contrast, the perpendicular component $\varepsilon_\perp$ is small, Fig.1(c). Thus the dielectric anisotropy $\Delta\varepsilon = \varepsilon_\parallel - \varepsilon_\perp$ is positive; it also fits well the Debye model with the same $\tau_D$. Upon cooling, both $\tau_D$ and the low-frequency $\Delta\varepsilon$ increase dramatically.

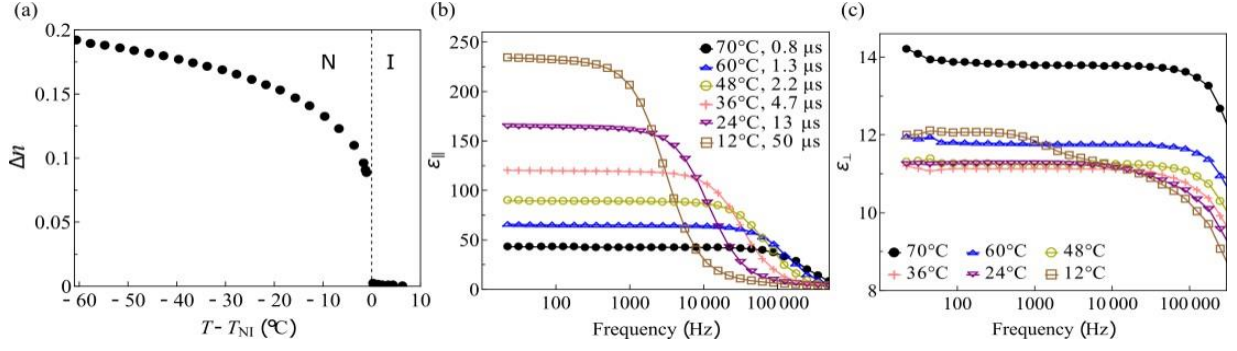

FIG. 1. Material parameters of GPDA200. (a) Temperature dependency of the birefringence at $\lambda = 632.8$ nm; (b) frequency dependencies of $\varepsilon_\parallel$ at various temperatures in the nematic phase and the corresponding Debye relaxation times, (c) $\varepsilon_\perp$ as a function of frequency at different temperatures in the nematic phase.

B. Experimental setup and method of electro-optic switching of NLC

To trigger the MEMOP effect in a material with $\Delta\varepsilon > 0$, the electric field should be applied parallel to the director $\hat{\mathbf{n}} = (n_x, n_y, n_z) = (0,0,1)$, which implies using a homeotropic cell. To maximize the field-induced birefringence, the light should propagate under some angle to the electric field; we chose the angle close to 45° by sandwiching the flat cell between two 45° prisms, Fig. 2(a). The homeotropic cell was composed of two parallel glass substrates with indium-tin oxide (ITO) electrodes of low resistivity ($10\,\Omega/\square$) and a polyimide alignment layer SE1211. The area of the electrodes was $2\times 2$ mm$^2$ and the cell thickness was in the range 4.4–7.0 μm.

A laser beam (He-Ne, $\lambda = 632.8$ nm) passed through the polarizer (P), LC cell, Soleil-Babinet compensator, and the analyzer (A) with the polarization direction being perpendicular to that of the polarizer, $A \perp P$. The transmitted light intensity was determined by using a photodetector TIA-525 (Terahertz Technologies) with the response time < 1 ns. The director $\hat{\mathbf{n}}$ makes a projection onto the plane formed by P and A; this projection makes an angle 45° with both P and A. A DC voltage pulse with the duration of 30 μs was applied using a waveform generator model 3390 (Keithley) and amplified by a wideband amplifier model 7602 (Krohn-Hite). The applied voltage pulses and photodetector signals were measured with 1G sample/s digital oscilloscope TDS2014 (Tektronix).

In order to measure the field-induced optical retardance $\delta\Gamma(t)$ and the corresponding effective birefringence $\delta n = \delta\Gamma(t)/d$, we used the four-point measurement scheme [13]. The technique is based on four successive measurements of the transmitted intensities $I_k$, ($k = 1, 2, 3, 4$) subjected to an identical electrical pulse at four different settings of the Soleil-Babinet compensator $\Gamma_k^{SB}$:

$$I_k(t) = [I_{max}(t) - I_{min}(t)]\sin^2\left\{\frac{\pi}{\lambda}\Gamma_k(t)\right\} + I_{min}(t) \qquad (2)$$

where $\Gamma_k(t) = \Gamma_{LC}(0) + \Gamma_k^{SB} + \delta\Gamma(t)$ is the total dynamic optical retardance of the system and $\Gamma_{LC}(0)$ is the optical retardance of the LC cell when no electric field applied. We select $\Gamma_k^{SB} = \lambda(m + k/4) - \Gamma_{LC}(0)$, where $m$ is an integer, in such way that the initial intensities with zero field are maximum $I_2(0)$, minimum $I_4(0)$ and mean $I_1(0) = I_3(0) = [I_2(0) + I_4(0)]/2$ values, Fig. 2(b), and for this selection the field-induced birefringence is calculated as

$$\delta n = \frac{\lambda}{2\pi d}\arg\left[I_2(t) - I_4(t) + i\left(I_1(t) - I_3(t)\right)\right], \qquad (3)$$

where $\lambda$ is the wavelength of the laser beam and $d$ is the thickness of the cell, Fig. 2(c). Note that Eq.(3) corrects the misprint in the corresponding equation in Ref. [13]. The switching-on time $\tau_{on}$ and the switching-off time $\tau_{off}$ are defined as the time when the field-induced birefringence changes between 10% and 90% of its maximum value, respectively.

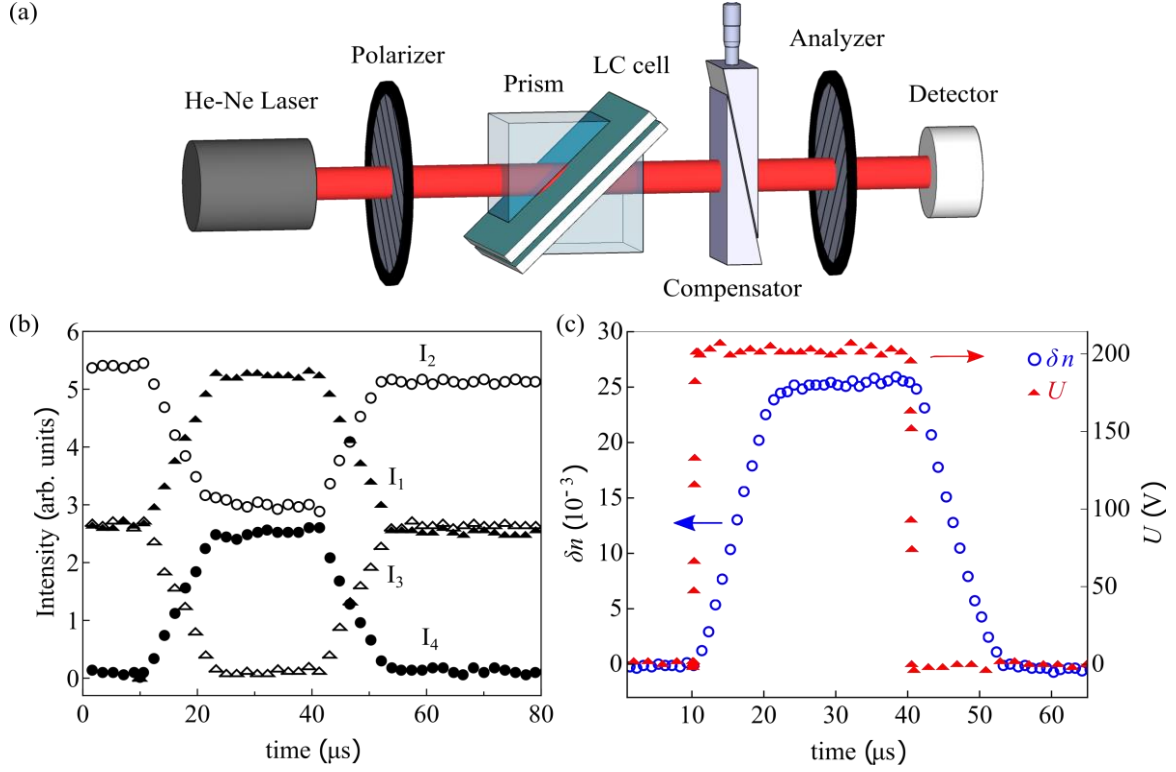

FIG. 2. Experimental methods. (a) Experimental setup. (b) Light intensity dynamics at four different compensator settings. (c) The dynamics of field induced birefringence $\delta n$ (circles) calculated from (b) and the corresponding voltage pulse $U$ (triangles).

## III. RESULTS AND DISCUSSIONS

We measure the electro-optical responses in the nematic phase in the temperature range $T = 28 - 72°C$. The microsecond dynamics of $\delta n$ at various applied electric fields and temperatures is shown in Figs 3(a,b). Higher electric fields and temperatures result in higher $\delta n$ values, Figs 3(a,b). The switching times $\tau_{on}$ and $\tau_{off}$ are on the order of microseconds, Figs 3(c,d). The saturation value of $\delta n$ shows a linear increase with an applied electric field, Fig. 3(e), and an increase with $T$, Fig. 3(f).

The field-induced birefringence $\delta n$ in the MEMOP effect in GPDA200 is significant and is achieved at electric fields that are by an order of magnitude lower than those reported previously for NEMOP in NLCs with a negative dielectric anisotropy [10-12]. For example, in the previously studied HNG715600-100 ( $\Delta\varepsilon = -12$, $T = 23°C$ ) $\delta n = 0.013$ was achieved under the applied field $E = 170$ V/μm [12], while the same $\delta n$ in GPDA200 is achieved already at $E$= (11-33) V/μm, depending on the temperature, Fig. 3(e).

The field-induced birefringence in materials with positive dielectric anisotropy is caused by the quenching of the director fluctuations [14] and modification of the uniaxial order parameter [11]. The modification of the uniaxial order parameter is proportional to the square of the field [11]. Quenching of the director fluctuations by the electric field is a well-known phenomenon associated with the anisotropy of the liquid crystals. For example, as shown by Dunmur et al, an electric field applied to a nematic with a positive dielectric anisotropy, tends to align the director parallel to itself and thus suppresses the director fluctuations [15]. Within the studied range of the applied electric fields, GPDA200 shows that the field-induced birefringence $\delta n$ grows linearly with the electric field, Fig. 3(e). The linear dependency indicates that the contribution to $\delta n$ from the director fluctuations quenching by the electric field dominates over the contribution from the enhanced uniaxial order parameter. This result could be explained by the fact that the weaker field ( $10^7$ V/m vs $10^8$ V/m ) still quenches director fluctuations through a strong dielectric coupling ( $\Delta\varepsilon = +200$ ), but at the same time, does not cause a strong modification of the uniaxial order parameter, which indicates a weak uniaxial susceptibility. Using the model [11], we calculate the field-induced birefringence caused by quenching of the director fluctuations in a homeotropic cell, as

$$\delta n = \frac{3\delta\tilde{\varepsilon}_f}{4\sqrt{2}n_e} \approx \frac{6k_B T \Delta n \sqrt{\varepsilon_0 \Delta\varepsilon} E}{\sqrt{2} K_{eff}^{3/2}} \, , \qquad (4)$$

where $\delta\tilde{\varepsilon}_f = \langle \tilde{\varepsilon}_{zz}(E) \rangle_V - \langle \tilde{\varepsilon}_{zz}(0) \rangle_V$ is the field induced modification of the optic tensor's $zz$ component averaged over the cell volume $V$, $n_e$ is the extraordinary refractive index of NLCs, $k_B = 1.38 \times 10^{-23}$ J/K is the Boltzmann constant, $\varepsilon_0 \approx 8.85 \times 10^{-12}$ F/m is the electric constant, and $K_{eff} = (4K_3)^{1/3} (K_1^{-1} + K_2^{-1})^{-2/3}$ is the effective elastic constant, $K_1$, $K_2$, and $K_3$ are the splay, twist, and bend elastic constants of NLCs, respectively.

It is of interest to compare the linear dependence of $\delta n$ on the electric field in our work with the field dependence of the induced birefringence $\delta n(E)$ reported by Dunmur, Waterworth, and Palffy-Muhoray for conventional nematics pentylcyanobiphenyl with $\Delta\varepsilon = 13$ [15], which is at least 10 times weaker than $\Delta\varepsilon$ of our material. Furthermore, in Ref. [15], the maximum applied electric field was 5 V/µm, which is about one order of magnitude smaller than the fields used in our experiments. Because of these large differences in the field and dielectric anisotropies, the data of Ref. [15] should be compared to the low-end of the field interval in Fig.3(e). A careful analysis demonstrate that in Ref. [15], the field-induced change of birefringence is in fact linear when the field becomes higher than $E_{fluct} = 0.3$ V/µm for a cell of thickness $d = 111$ µm, 0.6 V/µm for $d = 55$ µm, and 2 V/µm for $d = 15$ µm. These linear dependencies are similar to the linear dependincies observed in our work. At lower voltages, the dependence $\delta n(E)$ found in Ref. [15] is not linear and can be approximated as $\delta n(E) \sim E^2$. The reason is that the low fields quench only some of the fluctuations, those with the few lowest values of the normal component of the wavevector, $q_z = \frac{\pi}{d} m$, where $m = 1, 2, 3, \ldots$. As the field becomes stronger, it quenches larger and larger portion of the fluctuations spectrum and the dependence $\delta n(E)$ becomes linear, as evidenced both by our results in Fig. 3(e) and by the data in Ref. [15] for the fields higher than $E_{fluct}$ values presented above. Morphing of the dependence $\delta n \propto E^2$ into the linear one as the field increases above $E_{fluct}$ is supported by the thickness dependence $E_{fluct} = \beta / d\sqrt{\Delta\varepsilon}$, where $\beta \sim 100$ V is a fitting parameter, estimated from the analysis of data presented in Ref.[15]. For our material with $\Delta\varepsilon$

= 200 in a cell of thickness 6 µm, using the value $\beta \sim 100$ V, we find $E_{fluct} = \beta / d\sqrt{\Delta\varepsilon}$ to be about 1.2 V/µm, which is smaller than the weakest applied field used in our experiments, see Fig. 3(e). As a result, data shown in Fig.3(e) correspond to the range where the induced birefringence grows linearly with the electric field.

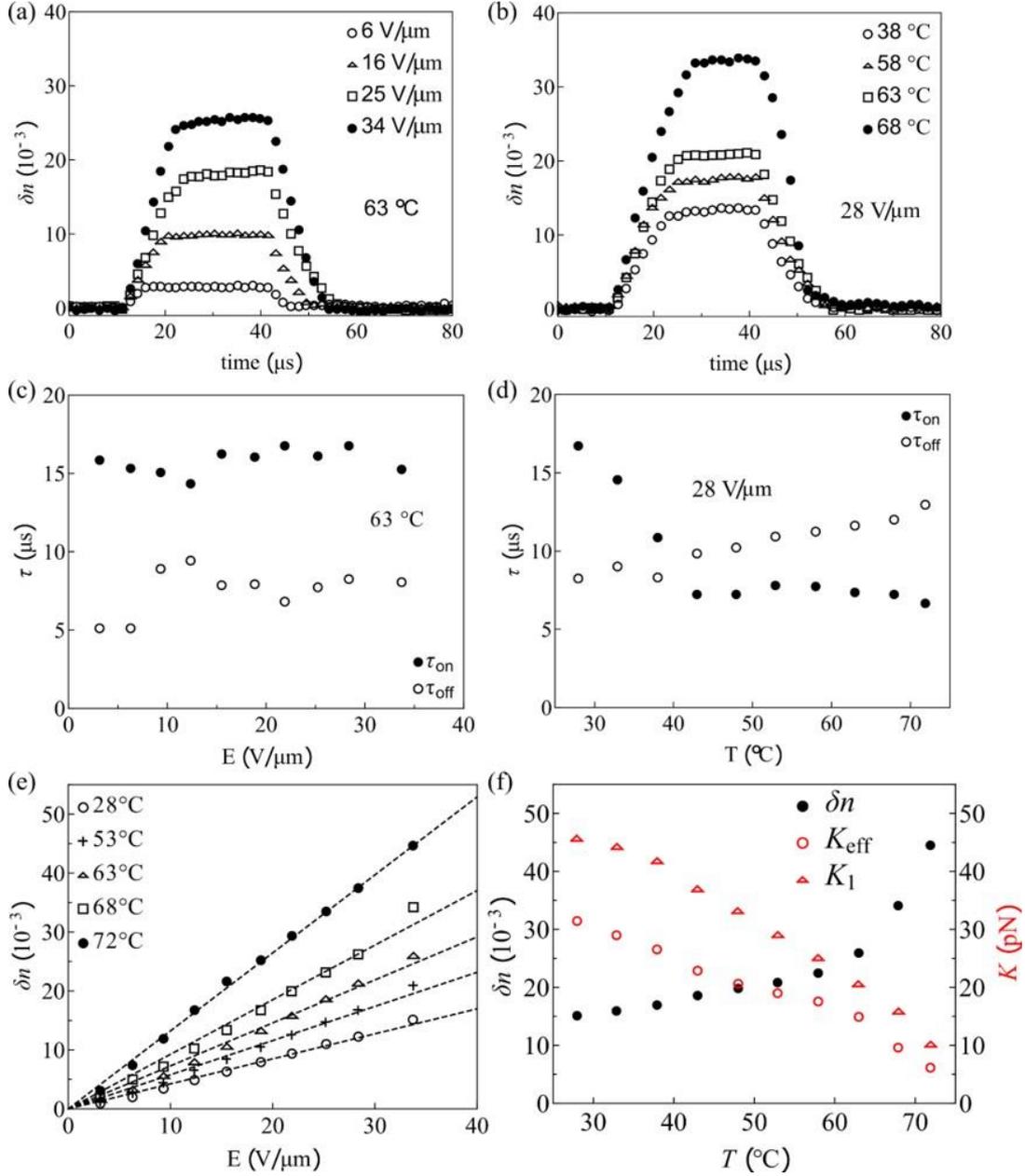

FIG. 3. Microsecond electro-optic switching of GPDA200 at various temperatures and different electric fields. (a) Dynamics of the field-induced birefringence at various electric fields at 63°C and (b) various temperatures at 28 V/µm. (c) Electric field dependence at 28 °C and (d)

temperature dependence at 28 V/µm of the characteristic response time. (e) Maximum value of the field-induced birefringence as a function of an electric field. The dashed lines show the linear fitting. (f) Temperature dependence of *δn* (34 V/µm), the effective elastic constant $K_{eff}$, and the splay elastic constant $K_1$ of GPDA200. $K_{eff}$ is calculated using Eq. (4).

The fluctuations are expected to grow as one approaches the clearing temperature, Eq. (4); this feature explains why the measured $\delta n$ increases with the temperature, Figs. 3(e,f). Furthermore, the director fluctuations, especially those with the small wave vector, should exhibit a somewhat slower relaxation. Using Eq. (4), we obtain the temperature dependence of the effective elastic constant $K_{eff}$, which has a similar temperature dependence as $K_1$ which we have measured directly using the splay Frederiks transition in a planar cell [1, 16, 17], see Fig. 3(f). The presented analysis demonstrates that the temperature dependence of the elastic constants prevails over the temperature dependence of the dielectric anisotropy; this prevalence explains the observed temperature dependence of *δn*, Fig.3(f).

The switching times $\tau_{on}$ and $\tau_{off}$ are almost independent on the applied electric field, Fig. 3(c), while their temperature dependences are different, Fig. 3(d). The monotonic increase of $\tau_{off}$ with temperature suggests that the ratio of a relevant viscous coefficient to the elastic constant is also increasing with the temperature. This result is rather surprising, since, usually, the viscous coefficient decreases with temperature much faster than the elastic constants. A plausible explanation might be in the multicomponent nature of the material and nano-segregation effects that produce different spectrum of fluctuation modes at different temperatures; however, further studies are needed to understand the effect better. In contrast, the temperature dependence of $\tau_{on}$ exhibits a crossover: at high temperatures, above 44 °C, $\tau_{on}$ remains almost constant, whereas below 44 °C, $\tau_{on}$ increases upon cooling. The dielectric torque depends on both the present and past values of the electric field and the director, when the rise time of the applied field is close to (or smaller than) the dielectric relaxation time [18, 19]. This is so-called dielectric memory effect. At the low temperature, Debye relaxation is slower than the rise of the applied voltage, Fig. 1(b). Thus, we attribute the low temperature increase of $\tau_{on}$ with the temperature dependency of the Debye relaxation time $\tau_D$, Fig. 1(b), and an associated memory effect in the field-induced polarization of the NLC. Figures 1(b) and 1(c) demonstrate that the Debye relaxation time is smaller than the pulse duration and therefore the memory effect [19] does not change the amplitude of electro-optical response.

Since the MEMOP response time $\sim 10\,\mu s$ is significantly shorter than that of the response time of the conventional electro-optic Frederiks transition, the resulting figure of merit $\text{FoM} = \delta\Gamma^2/\pi^2\tau_{off}$ of MEMOP is very high. In the Frederiks transition, a large $\tau_{off}$ represents the main bottleneck for applications. In the MEMOP effect, $\tau_{off}$ is practically the same as the field-on switching time $\tau_{on}$. With the typical values measured in our work, $\delta n = 0.04$ and $\tau_{off} = 10\,\mu s$, for a typical cell of a thickness $d = 5\,\mu m$, one finds $\text{FoM} \approx 10^2\,\mu m^2/s$. A typical Frederiks reorientation produces a weaker FoM~1-10 $\mu m^2/s$ [5]. Note here that the absolute limit of the induced birefringence in an NLC could not be higher than about $\delta n \sim 0.2$; the latter value corresponds to a perfect alignment of each and every molecule. However, the electric field required for creating such an idealistic perfectly aligned state with a maximum possible scalar order parameter $S = 1$, is much higher than the fields in our experiments. Consider, for example, a field with a dielectric extrapolation length that corresponds to a characteristic intermolecular distance $a$ of a few nanometers. Such a field $E_c = \pi\sqrt{K_{eff}/\varepsilon_0\Delta\varepsilon}/a$ corresponds to the limit of the continuous elastic theory and is approximately $E_c \approx 200\,V/\mu m$ for $a \approx 2\,nm$. The latter value is one order of magnitude larger than the fields in our experiments. Note, however that even with the field used in our work, one can easily achieve the levels $\delta\Gamma \geq \lambda/2$ of the field-induced optical retardance required for various electro-optical applications: With the typical $\delta n \approx 0.04$, the required condition $\delta\Gamma \geq \lambda/2$ is satisfied in cells of the thickness $10\,\mu\text{m}$ and above.

## IV. CONCLUSIONS

We demonstrated a MEMOP effect, a microsecond electro-optical response in the nematic material GPDA200 with an extraordinary high dielectric anisotropy $\Delta\varepsilon > +200$. The effect is mainly caused by the quenching of director fluctuations in the electric field that is applied parallel to the director. The field-induced birefringence in the range 0.01-0.04 is produced by electric fields on the order of $10^7$ V/μm. The described MEMOP effect yields the FoM one or two orders of magnitude higher than the conventional Frederiks effect. It is also of interest to compare the described MEMOP effect to the NEMOP effect observed in nematics with $\Delta\varepsilon < 0$ [10-12]. As compared to NEMOP, the MEMOP effect requires electric fields that are one order of magnitude lower to achieve a similar field-induced birefringence [12]. The trade-off between NEMOP and MEMOP is in the response time vs. the applied field. Although MEMOP is slower than NEMOP, its microsecond response times $\tau_{on} \approx \tau_{off} \approx 10\,\mu\text{s}$ are still very attractive for fast electro-optical applications. We conclude that the proposed MEMOP shows properties that fill the gap between the conventional relatively slow Frederiks effects and the ultrafast nanosecond responses of NEMOP. As such, MEMOP is promising for a variety of applications in devices such as electrically controlled shutters and beam steerers.

ACKNOWLEDGMENTS

We acknowledge useful discussions with Sathyanarayana Paladugu.